# Dimensionless Mapping: A Combinatorial Algorithm to Design Invisible Dopants

Mona Zebarjadi[1, 2], Wenqing Shen[1]

1-Department of Mechanical and Aerospace Engineering, Rutgers University

98 Brett Road, Piscataway, NJ, 08854, USA

2-Institute of Advanced Materials, Devices, and Nanotechnology, Rutgers University

607 Taylor Road, Piscataway, NJ 08854, USA

**Abstract**

Electronic cloaking has been recently suggested to design invisible dopants with electronic scattering cross sections smaller than 1% of the physical cross section ($\pi a^2$). Cloaking layers could be designed to coat nanoparticle dopants to minimally scatter conduction electrons and to enhance the electronic mobility. In some cases, such enhancements would result in larger thermoelectric power factors. The main difficulty is the fact that the created potential upon coating is not tunable and is determined by the band alignment of the chosen materials for the core, the shell and the host as well as the charge distribution in these layers. To find proper combinations of materials, one needs to probe a large class of materials combinations and layer sizes. This approach is time-consuming and impractical. Here we introduce a mapping method to identify possible combinations by comparing the dimensionless parameters of the chosen materials with the provided maps and without any transport calculations. Using this approach, we have identified several combinations of core, shell and host materials for which electronic cloaking is achievable. We have optimized the size and doping level of some of these materials combinations to maximize their thermoelectric power factor. Compared to traditional impurity-doped samples, up to 14.50 times improvement in the thermoelectric power factor was observed at T=77K.

## 1. INTRODUCTION

Design of objects invisible to electrons of specific energy is attractive for many applications such as electronic filters, switches, high mobility materials and fast response devices. [1–11] Manipulation of electron flow could also be achieved by near-zero refractive index semiconductor metamaterials [12] or by zero effective mass along certain direction [13]. Recently, it has been proposed that cloaking could be used to design invisible dopants to reduce the ionized impurity scattering rates and to enhance the electron mobility. [2] Core-shell structures that are following the shape of the Fermi surface are useful for such designs. For example when the Fermi surface is spherical (parabolic band structure), spherical core-shell particles could be used wherein the size and the band offsets are tuned to lower the electron-nanoparticle scattering cross section to values as low as 1% of the physical cross section, $\pi a^2$, in a narrow energy window (cloaking window). Here $a$ is the nanoparticle radius. When such nanoparticles are used as dopants, and when the cloaking window and the Fermi window are tuned to overlap, significant improvement in the electron mobility could be achieved compared to the conventional doping especially at low temperatures wherein impurity scattering is the dominant scattering. [14]

High mobility bulk materials are desired for many applications in the semiconductor device industry. One particularly interesting application is in the field of thermoelectrics. Good thermoelectric materials are required to have large electrical conductivity ($\sigma$), large Seebeck coefficient ($S$), and low thermal conductivity ($\kappa$). Invisible (electronically cloaked) dopants are suggested recently for such applications to minimize ionized impurity scattering and therefore to enhance the electrical conductivity. In addition, potentially, electronically cloaked dopants can lower the thermal conductivity since they are not invisible to phonons. For example, in the case of electronically cloaked core-shell nanoparticles, if the combination of core, shell and host materials is chosen to have large acoustic mismatch, they could scatter phonons significantly. This is not in conflict with nanoparticles being transparent to electrons. Therefore in principle materials with very low Lorenz numbers could be designed. Finally it is known that the Seebeck coefficient could be enhanced when there are sharp features in relaxation times, $\tau(\varepsilon)$, with respect to electron energy, $\varepsilon$. [15,16] Introducing cloaking windows could result in sharp features in the relaxation times, and therefore could enhance the Seebeck coefficient as discussed in an earlier publication. [14] Therefore, incorporation of invisible nanoparticle dopants could potentially improve all the three thermoelectric parameters simultaneously.

In a recent paper, we identified several combinations of hollow nanoparticles and host materials for which cloaking is possible. [17] Hollow nanoparticles are difficult to synthesize and therefore, there is a need to identify filled core-shell structures. In this work, we develop a new search method to identify embedded core-shell nanoparticles to improve mobility and thermoelectric properties of certain host materials.

The paper is organized in the following manner: first, we describe our methodology to calculate the scattering cross section. We then describe the developed search method, which we refer to as dimensionless mapping, to identify proper materials combinations for electronic cloaking. We use a small material database to test the feasibility of the proposed mapping method. Finally, two sets of identified materials combinations are optimized and the effect of nanoparticle doping on their thermoelectric properties is discussed.

## 2. METHODOLIGY

A core-shell structure is made of two co-centered step-function potentials. Different combinations of well-barrier, barrier-well, well-well or barrier-barrier could be used to observe cloaking states. One can theoretically tune the sizes, potentials and effective masses of the barriers and wells to minimize the electron-nanoparticle scattering cross section. The problem comes from the fact that in materials design, the potentials and effective masses are not tunable and they are set by the band-offset, the details of charge transfer between the three layers and the band structure of each layer.

This makes the search for proper combinations of the three materials, a time consuming and expensive task as there are too many parameters involved in the search. Even for the simplest parabolic band structure, conductivity effective mass, electron affinity, work function, the size of each layer, the temperature, the electron incident energy, and the nanoparticle volume fraction are the determining parameters for the electron-nanoparticle scattering rates. This makes the problem a function of 14 parameters and therefore one needs to search a huge parameter space to identify proper materials combinations. To do so, we break down these parameters.

First, we simplify the problem further to form a filtering process. We note that out of the 14 parameters involved in the calculation of electron-nanoparticle scattering rate, several (including the work functions in core, shell and host materials and temperature) could be eliminated, if we ignore charge transfer. Volume fraction is irrelevant if we discuss scattering off of one nanoparticle only and aim to design all the nanoparticles with the same geometry and materials composition. This processes lowers the number of parameters to 9 and could be used as the first filter to limit the number of materials combinations. Charge transfer could be included in the next round of optimization.

Now lets look at scattering cross section of an electron with incident energy, $\varepsilon$, from a spherical core-shell nanoparticle, ignoring charge transfer. We follow the same method as discussed in our previous publications [2,14] and use partial wave method [18] to calculate the total scattering cross section ($\Sigma$)

$$\frac{\Sigma}{\pi a^2} = \frac{4}{(ka)^2} \sum_{l=0}^{\infty} (2l+1) sin^2 \delta_l \tag{1}$$

$$\delta_l = \arctan\left(\frac{ka j_l'(ka) - \gamma_l a j_l(ka)}{ka n_l'(ka) - \gamma_l a n_l(ka)}\right) \tag{2}$$

$$\gamma_l a = \frac{m_h}{m_s} \frac{(\beta a)^2 j_l(\alpha a_c) n_l'(\beta a_c) j_l'(\beta a) - \frac{m_s}{m_c}\left[\alpha a \beta a j_l'(\beta a) j_l'(\alpha a_c) n_l(\beta a_c) + \alpha a \beta a j_l(\beta a_c) j_l'(\alpha a_c) n_l'(\beta a)\right] - \beta^2 a^2 j_l(\alpha a_c) j_l'(\beta a_c) n_l'(\beta a)}{\beta a j_l(\alpha a_c) n_l'(\beta a_c) j_l(\beta a) - \frac{m_s}{m_c}\left[\alpha a j_l(\beta a) j_l'(\alpha a_c) n_l(\beta a_c) + \alpha a j_l(\beta a_c) j_l'(\alpha a_c) n_l(\beta a)\right] - \beta a j_l(\alpha a_c) j_l'(\beta a_c) n_l(\beta a)} \tag{3}$$

We use the same terminology as used in **Ref. 2**. **Eqs. 1-3** are slightly modified to make them dimensionless. Here, $\delta_l$ is the phase shift of the $l^{\text{th}}$ partial wave, $k = \frac{\sqrt{2m_h\varepsilon}}{\hbar}$, $\alpha = \frac{\sqrt{2m_c(\varepsilon-V_c)}}{\hbar}$ and $\beta = \frac{\sqrt{2m_s(\varepsilon-V_s)}}{\hbar}$ are wave numbers in the host, core and shell regions respectively. $a_c$ is the core radius and $a$ is the nanoparticle radius. $V_c$ and $V_s$ are the potentials of the core and shell relative to the host respectively (i.e. the electron affinity differences of core and shell with respect to host). $m_h$, $m_c$ and $m_s$ are the effective masses of the host matrix, core and shell materials respectively.

By making the equations dimensionless, we note that the scattering cross section is only a function of 6 independent parameters namely $\alpha a, \beta a, ka, \frac{a_c}{a}, \frac{m_h}{m_s}$, and $\frac{m_s}{m_c}$. Instead of 9 parameters, now we only need to scan for six. Namely, instead of three masses, we have two mass ratios, instead of three electron affinities, we have two potentials (relative affinities) and instead of two radii, we have radius ratios. Next we separate the geometrical factors, which are irrelevant to materials scan. In our case, $\frac{a_c}{a}$ is the only geometrical factor (independent of the selection of materials). Finally, we separate the least sensitive parameters. These are the parameters versus which the scattering cross section has a smooth behavior without sudden extreme changes. In our case, the scattering rates are changing smoothly versus mass ratios. Therefore, we can use discretize points for the mass ratios and use interpolations if required in between the points. This process leaves us with three parameters $(\alpha a, \beta a, ka)$, which form a manageable 3D space for scanning. Note that such break down of parameters is not limited to the current problem and can be used in many different models. Even most complicated models with many parameters such as those used in biology could be broken down to 4 or 5 parameters to which the model is most sensitive. [19]

For each different discrete mass ratio, we prepare a $(\alpha a, \beta a, ka)$ map. The dimensionless maps are 3D spaces where in the axes are $\alpha a, \beta a, and\ ka$. For each point in this space (i.e. a given $\alpha a, \beta a$ and $ka$), we scan $\frac{a_c}{a}$ ratio and calculate the scattering cross section using **Eqs. 1-3**. We accept the combination of $\alpha a, \beta a$, and $ka$ if the relative scattering cross section $\frac{\Sigma}{\pi a^2}$ is less than 0.01 (1%) for any reasonable $\frac{a_c}{a}$ ratio and we place a dot in the accepted $(\alpha a, \beta a, ka)$ coordinate. We also save the value of $\frac{a_c}{a}$ for later use.

After developing the maps, for a given combination of real materials for core, shell and host, we select a map with the closest mass ratios. Next we decide on the incident electron energy. Since the idea is to overlap the cloaking window and the Fermi window, the electron energy should be chosen to be within $K_BT$ of the optimum Fermi level. The optimum Fermi level is the Fermi level at which the thermoelectric power factor of the host material (conventionally doped) is maximum. At this stage, with a good certainty and simply by positioning the dimensionless parameters of the materials combination in the map and identifying the overlapping regions, we can decide whether or not the chosen combination will have cloaking windows! As soon as we decide on the electron energy, $\alpha, \beta$, and $k$ are known and therefore $(\alpha a, \beta a, ka)$ will be a line in the map (with varying $a$). If the line crosses the saved matrix points, then there is chance of observing cloaking states and if not, cloaking is unlikely.

## 3. RESULTS AND DISCUSSIONS

Six mass ratio values (0.5, 1, 1.5, 2, 3, 5) of $\frac{m_s}{m_c}$ and $\frac{m_h}{m_s}$ are considered to form 36 sets of maps in total. **Fig. 1** demonstrates the two dimensional projection of some of the maps created for the mass ratios of 1 and 5. The values of $\alpha$ *and* $\beta$ are either pure real or pure imaginary. The imaginary part in the map indicates that the energy of the electron is lower than the potential of the corresponding layer and therefore the electrons are tunneling.

An interesting observation from these maps is that even in the case of barrier-barrier two-step potential, cloaking is possible. However in most cases, the corresponding $ka$ values are small ($ka < 0.5$). Another remark is to note that for certain set of parameters, the cloaking is achievable regardless of the value of $\alpha a$. For example, in **Fig. 1(b)**, when $ka$ is around 2 and $\beta a$ is around 1, cloaking is achievable for any arbitrary value of $\alpha a$. This makes the choice of the proper material for the core an easy choice. In such cases, the only relevant parameter for the host is its effective mass. Finally, an exciting observation is that even for relatively large $ka$ values ($ka \sim 5$), cloaking is possible. This is very important since it indicates that one can observe cloaking for large size particles. In fact we observed that for some of the identified combinations, the radius of invisible nanoparticles could reach to 10nm. Note that the cancellation of the first two partial waves technique developed in our first publication relied on the fact that higher order partial waves could be ignored only when $ka$ is less than or equal to 1. [2] Such restriction limits the size of cloaked nanoparticles to small values (couple of nanometers). Cloaking could be achieved when the $0^{th}$ order partial wave gives nearly zero scattering cross section and the contribution from the higher-order partial waves is relatively small, as shown in **Fig. 2(a)** and **(c)**. In these examples, the minima of the scattering cross section of the $0^{th}$ and $1^{st}$ order partial waves do not overlap. However, at the minimum of the $0^{th}$ one, the contributions from higher order partial waves are negligible, making the minimum of the total scattering cross section reaching values less than 1% of the physical cross section. Most of the larger observed $ka$ values show this feature in their scattering rates.

We conclude that to achieve extremely small scattering cross sections, none of the previously thought conditions are necessary. In other words, the co-existence of barrier and well, the overlap of the first two minima, and small $ka$ values (less than or close to one) are not required and the phase-space of observing cloaking is much larger than what was believed previously.

To test the feasibility of the dimensionless mapping method, we form a materials database (**Table I**). The database includes 15 materials. These materials could be used for core-shell-host combinations. Combinations of A-A-A are not allowed but A-B-A combinations are allowed. More over for each set of A-B-C or A-B-A we can use n-type or p-type doping. Therefore, there will be 15×14×15×2=6300 different combinations to test. Maps with closest mass ratios are adopted for each materials combination. In this process, several selection rules were enforced. First, only nanoparticles with core radius larger than 1 nm and shell thickness larger than 1 nm were accepted. Second, only adjacent layers with lattice mismatch less than 5% were allowed. Finally, for each materials combination, a proper energy range was identified for the host matrix to be close to the optimum Fermi level. Then intersection of the $(\alpha a, \beta a, ka)$ lines and the relevant maps were evaluated to identify cloaking points. Using the dimensionless mapping method, the original 6300 combinations were narrowed down to 14 combinations, which satisfied all of the search criteria enforced. These fourteen good combinations are listed in **Table II**. All of the 14 identified combinations were tested later with their exact mass values. All combinations could satisfy the size criteria and have minimal relative scattering cross section less than 3% and majority of them could reach below 1%. Note that this is not guaranteed in the first filtering processes since only maps with the closest mass ratios (and not the exact mass ratios) are chosen and therefore mass ratios in the first scan are not exact. Also note that one cannot afford to form an infinite number of maps with all possible mass ratios. Therefore, there is a tradeoff between the accuracy of the filtering and the cost of the calculations. The main advantage of the method is that the maps could be calculated once and saved as a database for permanent use.

The main purpose of the nanoparticles in the context of thermoelectric materials is to dope the host matrix. Therefore the effect of charge transfer from nanoparticles to the host matrix cannot be ignored. In the next step, we need to calculate the effect of charge transfer in the identified combinations. Finally optimization of the nanoparticle size, doping level and volume fraction is required to optimize the thermoelectric properties of the designed core-shell nanoparticle doped host materials.

We have performed partial optimization for two of the identified combinations, namely combination 1: GaAs, InP, Ge and 2: $Ga_{0.47}In_{0.53}As$, InAs, InP in the order of host, core and shell. The total scattering cross section (with and without charge transfer) of the optimized nanoparticle doped matrix made out of combination 1 is shown in **Fig. 2(a)** and that of combination 2 is shown in **Fig. 2(c)**. In both tested cases, the net effect of considering charge transfer is to shift the cloaking window to smaller energies.

For the optimized nanoparticle size, the Fermi level is modified by changing the nanoparticle doping density. The thermoelectric power factor ($\sigma S^2$) is then calculated for each Fermi level. To do so, we first calculate the nanoparticle momentum relaxation time ($\tau_m$):

$$\Sigma_{\mathrm{m}}(\varepsilon) = \frac{4\pi}{k^2}\left[\sum_{l=0}^{\infty}(2l+1)\sin^2\delta_l - \sum_{l=0}^{\infty} 2l\cos(\delta_l - \delta_{l-1})\ \sin\delta_l \sin\delta_{l-1}\right] \tag{4}$$

$$\frac{1}{\tau_m(\varepsilon)} = \Sigma_m(\varepsilon) v_g(\varepsilon)\rho_{np} \tag{5}$$

$\Sigma_m$ is the momentum scattering cross section, $\rho_{np}$ is the nanoparticle density and $v_g$ is the group velocity. We use the Matthiessen's rule to add the nanoparticle momentum scattering rates ($\frac{1}{\tau_m}$) to other relevant relaxation rates (phonons, alloy, etc.) and calculate the total relaxation rate ($\tau(\varepsilon)$). Transport properties then could be calculated using

$$\sigma = e^2 \int_{-\infty}^{+\infty} d\varepsilon \left(-\frac{\partial f_0}{\partial \varepsilon}\right) \sigma(\varepsilon) \tag{6}$$

$$S = \frac{1}{Te}\left[\frac{\int_{-\infty}^{+\infty} d\varepsilon \left(-\frac{\partial f_0}{\partial \varepsilon}\right)\sigma(\varepsilon)(\varepsilon-\varepsilon_f)}{\int_{-\infty}^{+\infty} d\varepsilon \left(-\frac{\partial f_0}{\partial \varepsilon}\right)\sigma(\varepsilon)}\right] \tag{7}$$

$$\sigma(\varepsilon) = g(\varepsilon) v_g(\varepsilon)^2 \tau(\varepsilon) \tag{8}$$

Here $\sigma(\varepsilon)$ is the differential conductivity, $g(\varepsilon)$ is the density of states, and $f_0$ is the Fermi Dirac function. Finally the calculated properties are compared to those of conventional impurity doping.

For combination 1 (GaAs host matrix), the electrical conductivity could be greatly improved by using nanoparticle dopants at Fermi levels deep inside the conduction band, corresponding to large doping densities (see **Fig. 3(a)**). At low doping densities, the scattering rate is dominated by other scattering mechanism, namely background phonons. By using core-shell nanoparticle doping, the Seebeck coefficient decreases slightly. The net improvement in the thermoelectric power factor is about 1450%, which is the result of improved mobility (see **Fig. 3(c)**).

Combination 2 ($Ga_{0.47}In_{0.53}As$ host matrix) is also tested and optimized. The optimized potential profile with and without charge transfer is shown in **Fig. 2(d)**. **Fig. 2(c)** shows that after considering charge transfer, the minimum scattering cross section is smaller and shifts to lower energies. **Fig. 4** shows the thermoelectric properties comparison between nanoparticle doping and uniform impurity doping. Similar to what was observed for combination 1, the electrical conductivity is increased and the Seebeck coefficient is slightly decreased. The maximum power factor is improved by 18% by using nanoparticle doping. The improvement of electrical conductivity is much lower compared to combination 1. This is because alloy scattering is quite large in $Ga_{0.47}In_{0.53}As$ and lowers the effect of improved doping, while there is no alloy scattering in the host material (GaAs) of combination 1.

The reduction in the Seebeck coefficient in both cases could be explained by the enhancement in the electrical conductivity, which increases the denominator of **Eq. 7**. **Fig. 5** compares electronic band structure contribution ($h(\varepsilon) = e^2\left(-\frac{\partial f_0}{\partial \varepsilon}\right) g(\varepsilon) v_g(\varepsilon)^2$), the relaxation times ($\tau(\varepsilon)$), and the differential conductivity ($h(\varepsilon)\tau(\varepsilon) = e^2\left(-\frac{\partial f_0}{\partial \varepsilon}\right)\sigma(\varepsilon)$) for $Ga_{0.47}In_{0.53}As$ host matrix doped with invisible dopants (combination 2) and conventional dopants (single impurity). In both cases the carrier density is $n = 2\times10^{22} m^{-3}$ corresponding to a Fermi level of -0.52 meV. As could be seen in **Fig. 5(a)**, by using invisible dopants, the slope of the relaxation times increases appreciably, however, its absolute value also increases significantly. This results in increase of the electrical conductivity and reduction of the Seebeck coefficient. This is also reflected in **Fig. 5(c)**. Another explanation is the fact that the peak value of $h(\varepsilon)\tau(\varepsilon) = e^2\left(-\frac{\partial f_0}{\partial \varepsilon}\right)\sigma(\varepsilon)$ shifts to energies closer to the Fermi level, resulting in lower Seebeck coefficients. [20] Enhancement in the Seebeck coefficient is possible for cloaking windows designed at higher energies. [14]

## 4. CONCLUSIONS

The dimensionless mapping strategy described could be applied to other blind combinatorial materials searches wherein many parameters are involved in the scan and could provide an accessible method to design new optical, acoustic and electronic meta-materials. In this work, we used the proposed mapping method to identify possible combinations of core-shell-host materials to achieve electronic cloaking. An extremely expensive task of searching a 14 dimensional space is reduced down to a simple table-look up task. A set of maps based on dimensionless parameters was formed to screen out good materials combinations for electronic cloaking. We formed 6300 materials combinations and narrowed them down to 14 optimistic combinations by simply comparing their dimensionless parameters with the provided maps. Two of these combinations were optimized with respect to their nanoparticle sizes and their doping density. We showed that by using invisible nanoparticle doping instead of conventional uniform impurity doping, the electrical conductivity and the power factor could be significantly improved. 14 times and 18% improvement in the power factor was reported for GaAs and InGaAs respectively at T=77K. Using the developed dimensionless maps, we showed that the phase space for achieving electronic cloaking is much larger than what was previously thought. In particular, many combinations of materials are possible to achieve electronic cloaking and it is possible to design large size invisible nanoparticles (radius > 10 nm).

**Acknowledgements**

This work is supported by the Air Force's Young Investigator Research grant No. FA9550-14-1-0316.

TABLE I. Material data base (Most of data @300K)

| Name | Hole Effective Mass ($m_0$) | Electron Effective Mass ($m_0$) | Eg (eV) | Electron Affinity (eV) | Lattice Constant (Å) |
|---|---|---|---|---|---|
| Ge [21] | 0.33 | 0.119 | 0.66 | 4.0 | 5.658 |
| Si [21] | 0.49 | 0.26 | 1.12 | 4.05 | 5.431 |
| GaAs [21] | 0.51 | 0.063 | 1.424 | 4.07 | 5.65325 |
| GaP [21] | 0.79 | 0.3 | 2.26 | 3.8 | 5.4505 |
| InAs [21] | 0.41 | 0.023 | 0.354 | 4.9 | 6.0583 |
| InP [21] | 0.6 | 0.08 | 1.344 | 4.38 | 5.8687 |
| AlAs [21] | 0.75 [22] | 0.262 [22] | 2.168 | 3.5 | 5.6611 |
| GaN (β) [21] | 1.3 | 0.13 | 3.2 | 4.1 | 4.52 |
| GaN (α) [21] | 1.4 | 0.2 | 3.39 | 4.1 | 5.186 |
| $Ga_{0.51}In_{0.49}P$ [23] | 0.7 | 0.088 | 1.849 | 4.1 | 5.653 |
| $Ga_{0.47}In_{0.53}As$ [21] | 0.45 | 0.041 | 0.74 | 4.5 | 5.8687 |
| PbTe | 0.024@4K [24] | 0.035@4K [24] | 0.3 [25] | 4.6 [25] | 6.462 [26] |
| PbS | 0.083 [24] | 0.087 [24] | 0.42 [26] | 4.6 [27] | 5.936 [26] |
| CdTe [28] | 0.8 | 0.09 | 1.49 | 4.28 [29] | 6.482 |
| ZnTe | 0.68 [30] | 0.12 [31] | 2.35 [26] | 3.53 [29] | 6.1 [28] |

TABLE II. Good Combination

| Index | Type | Host | Core | Shell |
|---|---|---|---|---|
| 1 | n | GaAs | InP | Ge |
| 2 | n | $Ga_{0.47}In_{0.53}As$ | InAs | InP |
| 3 | n | GaAs | PbS | Ge |
| 4 | n | GaN(α) | Ge | Si |
| 5 | n | GaN(α) | GaN(α) | Si |
| 6 | n | GaN(α) | $Ga_{0.51}In_{0.49}P$ | Si |
| 7 | n | $Ga_{0.51}In_{0.49}P$ | InP | Ge |
| 8 | n | $Ga_{0.51}In_{0.49}P$ | $Ga_{0.47}In_{0.53}As$ | Ge |
| 9 | n | $Ga_{0.51}In_{0.49}P$ | PbS | Ge |
| 10 | n | $Ga_{0.51}In_{0.49}P$ | GaN(α) | Si |
| 11 | n | GaAs | $Ga_{0.47}In_{0.53}As$ | Ge |
| 12 | n | $Ga_{0.47}In_{0.53}As$ | PbS | InP |
| 13 | p | InAs | InAs | $Ga_{0.47}In_{0.53}As$ |
| 14 | p | $Ga_{0.47}In_{0.53}As$ | $Ga_{0.47}In_{0.53}As$ | InAs |

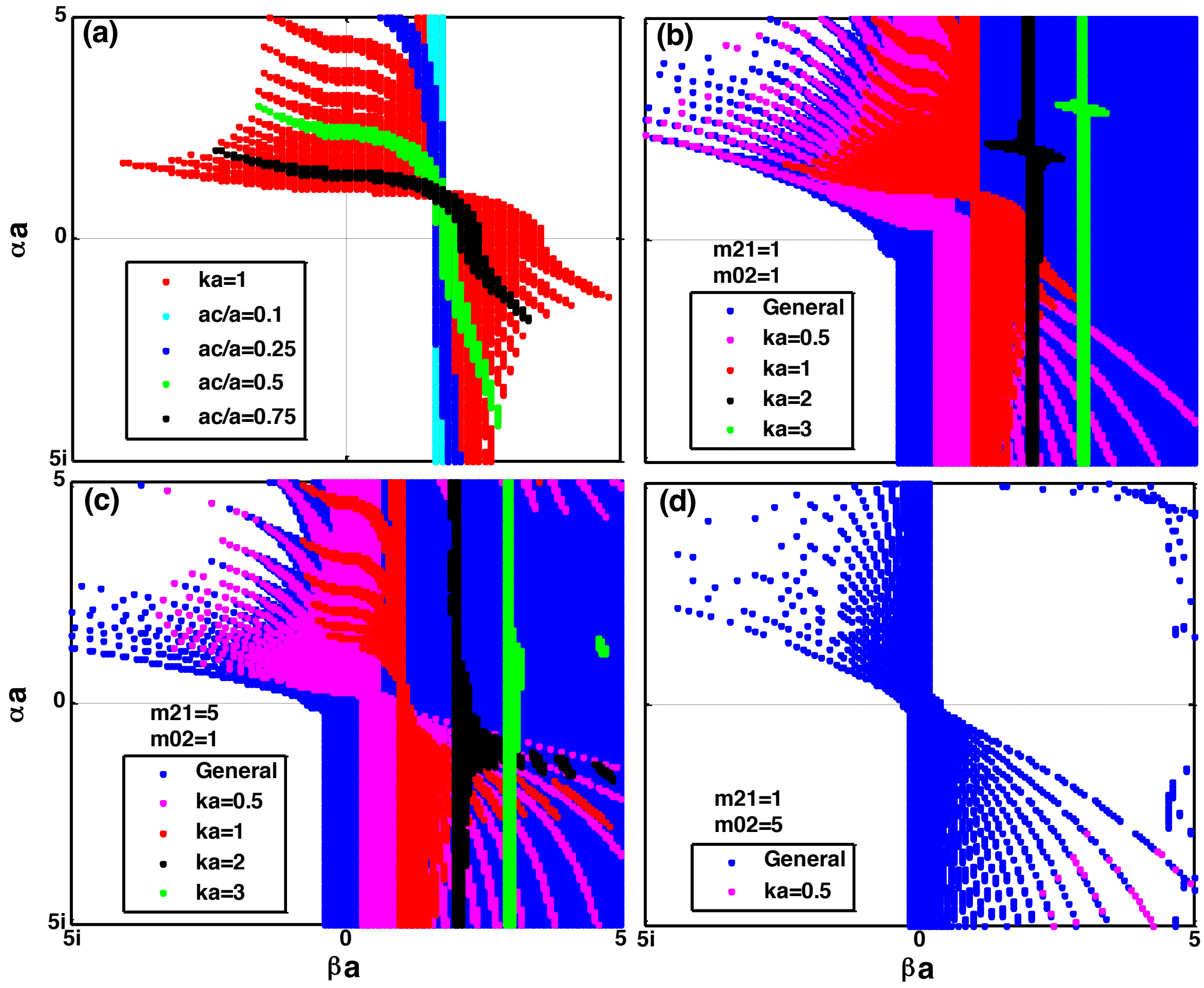

**Fig. 1** (color online) General dimensionless maps: $\beta a$ and $\alpha a$ solutions for which the total scattering cross section of the electron-nanoparticle is smaller than 1% of the physical cross section. $\beta a$ and $\alpha a$ are either pure real or pure imaginary, both cases are plotted here. a) $\beta a$ and $\alpha a$ solutions for $ka = 1$ and $\frac{m_h}{m_s} = \frac{m_s}{m_c} = 1$, are plotted using red dots. This plot is then broken down to the corresponding $\frac{a_c}{a}$ values. For example black color refers to obtained solutions when $\frac{a_c}{a}$ ratio is set equal to 0.75, and so on. Other plots: Blue dots represent the general map for (b) $\frac{m_h}{m_s}$, and $\frac{m_c}{m_s} = 1$, (c) $\frac{m_h}{m_s} = 1$ and $\frac{m_s}{m_c} = 5$, and (d) $\frac{m_h}{m_s} = 5$, $\frac{m_s}{m_c} = 1$. In these plots all $ka$ values are scanned. These are in fact 3D plots wherein $ka$ is the z-axis. Here we simply show the projection on a 2D plot. The solutions are then broken down to different planes corresponding to different $ka$ values as indicated in the legend. In (d), solutions for $ka$ = 1, 2 and 3 are not plotted as there is no satisfied point in those regions.

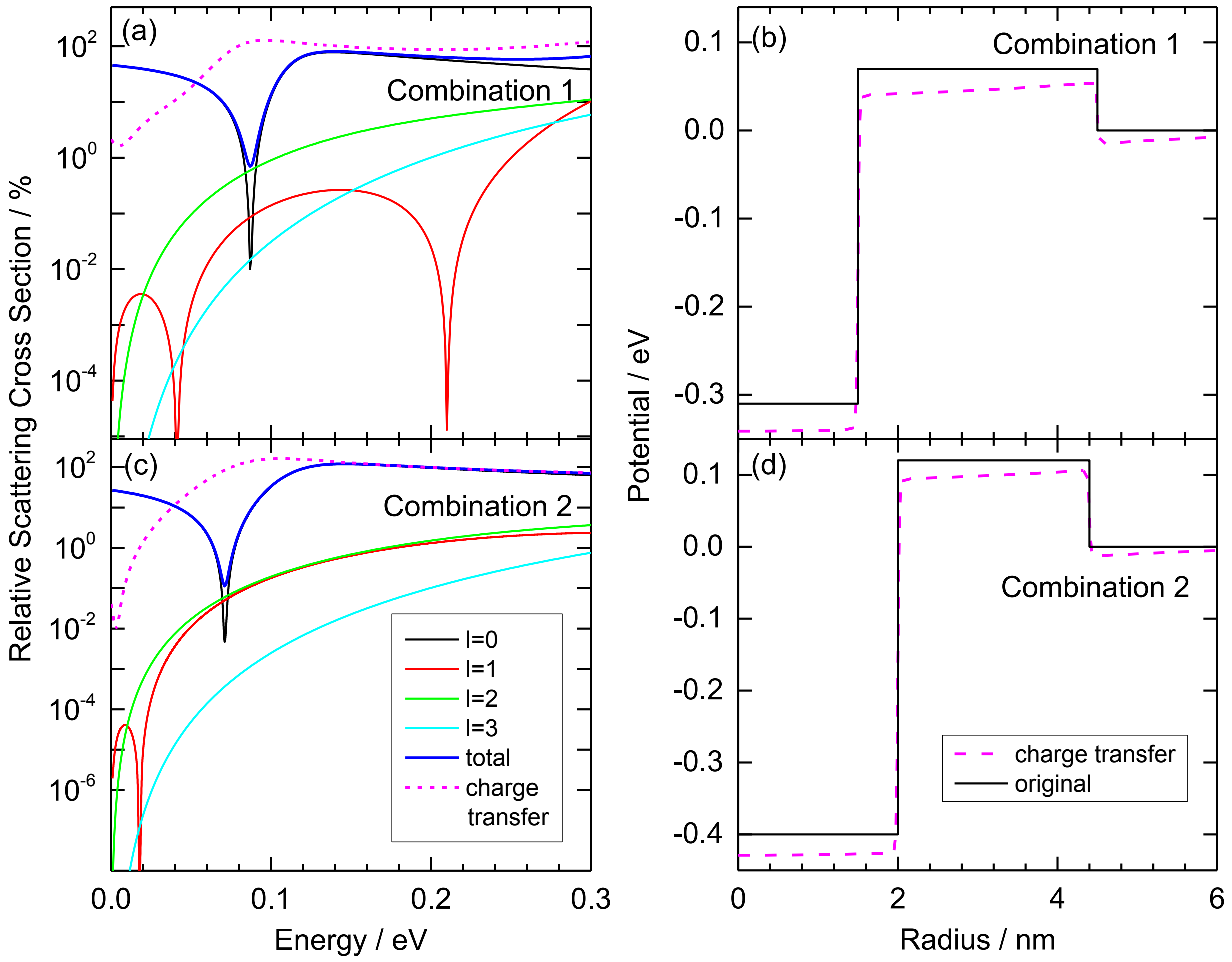

**FIG. 2** (color online) a,c) The total scattering cross section relative to the physical scattering cross section and the contribution from the first 4 partial waves without considering charge transfer (Solid lines). The dashed line shows the total scattering cross section after considering charge transfer. c,d) Nanoparticle potentials: Black solid lines show the 2-step potential without considering the charge transfer and the magenta dashed ones show the potential with consideration of charge transfer. Parameters for (a, b) calculated for combination 1 (GaAs-InP-Ge): The radius of nanoparticle is $a = 4.5\ nm$, and the radius of core is $a_c = 1.5\ nm$. When charge transfer is accounted for, the doping density is $3\times10^{22}\ m^{-3}$corresponding to 1 electron per nanoparticle. Parameters for (b, d) calculated for combination 2 ($Ga_{0.47}In_{0.53}As$- InAs- InP): The radius of nanoparticle is $a = 4.4\ nm$, and the radius of core is $a_c = 2\ nm$ . When charge transfer is accounted for, the doping density is n=$2\times10^{22} m^{-3}$corresponding to 1 electron per nanoparticle.

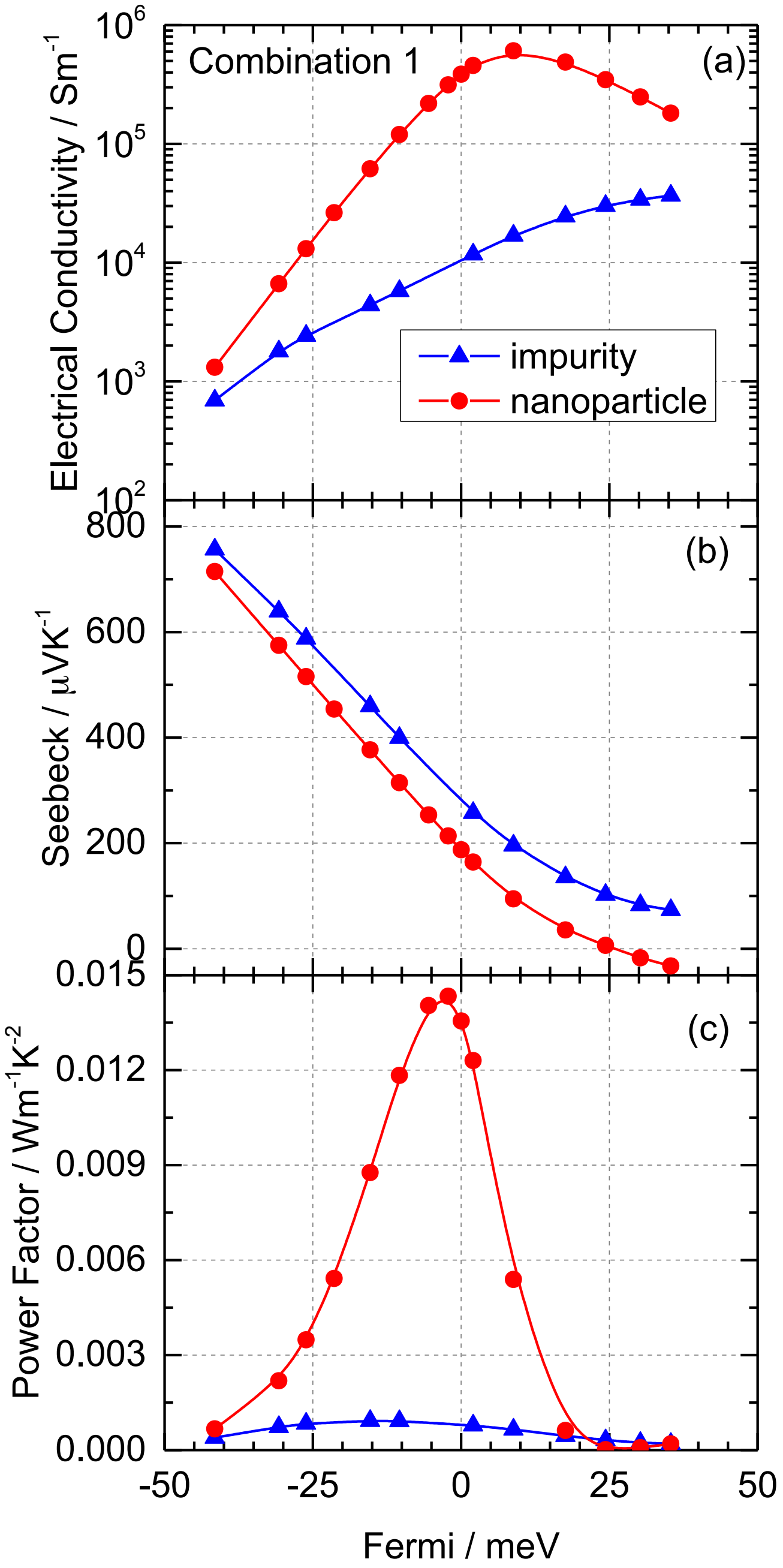

**Fig. 3** (color online) Comparison of electrical conductivity, Seebeck coefficient and power factor for different doping methods. The blue curve is GaAs doped with invisible nanoparticle dopants (combination 1 in Table II). The geometry of the designed nanoparticles is described in **Fig. 2**. The red curve labeled by 'impurity' refers to conventional uniform impurity doping of the host matrix (GaAs).

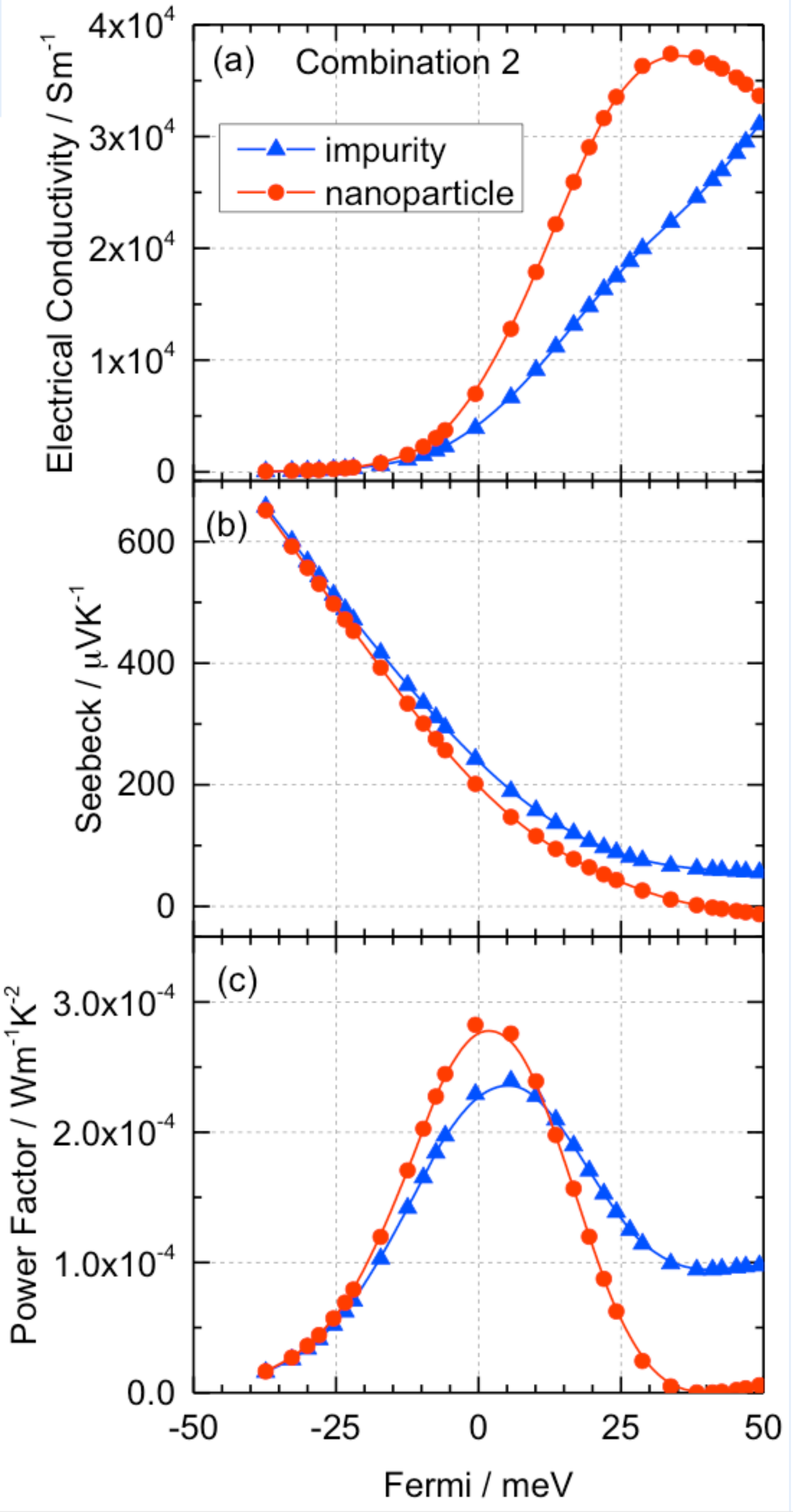

**Fig. 4** (color online) Comparison of electrical conductivity, Seebeck and power factor for different doping method. The blue curves is for combination 2 in **Table II** and the geometry is described in **Fig. 2**. The curve labeled by 'impurity' refers to conventional uniform impurity doping and the curve labeled by 'nanoparticle' means material doped by core-shell nanoparticles.

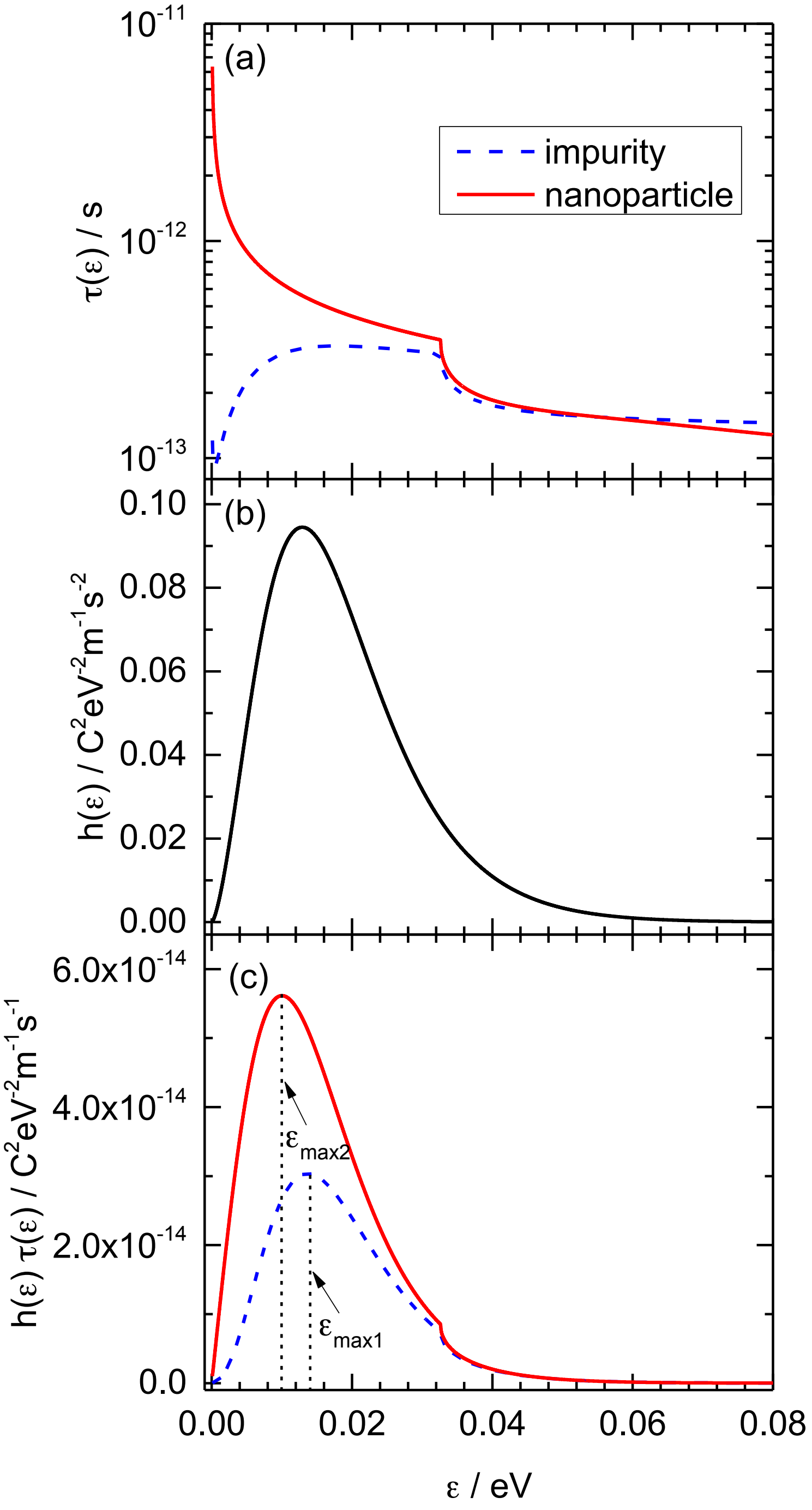

**Fig. 5** (color online) a) Comparison of $\tau(\varepsilon)$; b) Curve of h($\varepsilon$); c) Curve of $h(\varepsilon)\tau(\varepsilon)$. The curve labeled by 'impurity' refers to conventional impurity doping and the curve labeled by 'nanoparticle' means material doped by core-shell nanoparticles.